\documentclass{appolb}
\usepackage{graphicx}
\usepackage{cite}
\newcommand{\wasa}{\mbox{WASA-at-COSY}}
\begin{document}
\title{Search for the manifestation of the mesic-nuclei on the $dd \rightarrow  {{^3}\mbox{He}} N \pi$ excitation function measured with \wasa 
\thanks{Presented at International Symposium on "Mesic Nuclei", Krak\'{o}w, Sept. 20, 2013}
}
\author{Wojciech Krzemien$^1$, P. Moskal$^{1,2}$ and M. Skurzok$^1$ for the \wasa~collaboration
\address{$^1$M. Smoluchowski Institute of Physics, Jagiellonian University, 30-059 Cracow, Poland}
\address{$^2$IKP, Forschungszentrum J\"ulich, D-52425 J\"ulich, Germany}
}
\maketitle
\begin{abstract}
We search for an evidence of $\eta$-mesic ${^4\mbox{He}}$ with the \wasa~detector.
Two dedicated experiments were performed at the Cooler Synchrotron COSY-J\"ulich.
The experimental method is based on the measurement of the excitation functions for the two reaction channels: $dd \rightarrow {^3\mbox{He}} p \pi^{-}$ and $dd \rightarrow {^3\mbox{He}} n \pi^{0}$, where the outgoing $N-\pi$ pairs originate from the conversion of the $\eta$ meson
on a nucleon inside the ${\mbox{He}}$ nucleus. In this contribution, the experimental method is shortly described  and  preliminary excitation functions from the 2010 data are presented.
\end{abstract}
  
\section{Introduction}
The studies of the nucleus systems in which one or more nucleon are replaced by another particle have been proven as an important tool of investigations in many different topics in the nuclear and in particle physics. 
E.g. the experiments on the hypernuclei opened a new area of strangeness physics. More recently, the studies of meson-nucleus bound states attracted many attention because they are treated as an excellent source of information about the meson properties in the nuclear matter, which is directly linked to the underling symmetries and structure of the QCD vacuum~\cite{Itahashi,Mesic-Bass, EtaMesic-Hirenzaki}. 
Many different systems have been studied e.g. pionic atoms~\cite{Itahashi}, kaonic systems~\cite{Silarski,Sakuma,Ishiwatari},  $\eta'$ and $\eta$ nuclei~\cite{Gal,wilkin2,bass,eta-prime-mesic-Nagahiro,Mesic-Bass,EtaMesic-Hirenzaki,eta-prime-mesic-Hirenzaki,eta-prime-mesic-Nagahiro-Oset,
ETA-Friedman-Gal,ETA-Gal-Cieply,Wycech-Acta,WYCECHGREE-GW-zGAla,Mesic-Kelkar,Bass10Acta, eta-prime-mesic-GSI-tanaka,lampf,lpi,gsi,gem,jurekmeson08, Magda2, moskalsymposium,jurek-he3,timo,meson08, fujiokasymposium, mami2, jinr,cbelsa, Adlarson2013,LPI, ELSA-MAMI-plan-Krusche}.            
The search for the $\eta$-mesic bound state has been and is being performed by many experiments,
 but so far no firm experimental confirmation of the existence of mesic nuclei has been achieved.

The observation of a steep rise in the total production cross-section and the phase variation of the scattering amplitude in the close-to-threshold region in the $dd \rightarrow ^4$He$ \eta$ reaction are interpreted as possible indications of a $^4$He$-\eta$ bound state~\cite{wilkin2}. 

\section{Method and previous results}

\begin{figure}[htb]
\centerline{%
      \scalebox{0.2}
         {
              \includegraphics{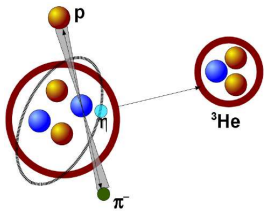}
         }
}
\caption[Bound state decay scheme]{\label{decay_scheme_p}Schematic picture of the  $({^4\mbox{He}}-\eta)_{bound} \rightarrow {^3\mbox{He}} p \pi^{-}$
decay. In the first step the $\eta $ meson is absorbed on one of the neutrons and the $N^{\star} $ resonance is formed. Next, the N$^{\star}$ decays into a $p-\pi^{-}$ pair. The $^3\mbox{He}$ is a spectator. Adapted from~\cite{Adlarson2013}. }
\end{figure}

We perform the search of a $\eta$-mesic helium produced in deuteron-deuteron collisions.
One of the possible decay scenarios of the $\eta-{^4\mbox{He}}$ bound state is 
the absorption of the $\eta$ meson on one of the 
nucleons in the $^4$He nucleus, leading to the excitation of the  $N^{*}$ (1535)
resonance which subsequently decays in a pion-nucleon pair.
The remaining three nucleons are spectators forming  a $^3$He
or $^3$H nucleus.
This scenario is schematically presented in the Fig.~\ref{decay_scheme_p}.

\begin{figure}[htb]
\centerline{%
      \scalebox{0.3}
         {
              \includegraphics{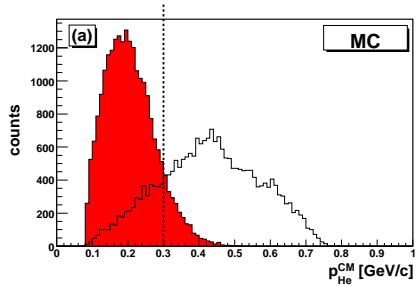}
         }
}
\caption[Distribution of the $^3$He momentum in c.m.]{\label{pmomc.m._p}Distribution of the $^3$He momentum in 
the c.m. system simulated for the processes leading to the creation
of the ${^4\mbox{He}}-\eta$  bound state:
$dd \rightarrow (^4{\mbox{He}} \eta)_{bound} \rightarrow ^3$He$p\pi^{-}$ (red area)
and of the phase-space $dd \rightarrow ^3$He$ p\pi^{-}$ reaction (black line).
Adapted from~\cite{Adlarson2013}.}
\end{figure}

The outgoing $^3\mbox{He}$ nucleus is a spectator and, therefore,
we expect that its momentum in the center of mass (c.m.) frame is relatively low and can be approximated 
by the Fermi momentum distribution of the nucleons inside the $^4\mbox{He}$ nucleus.
This signature allows us to suppress the background from reactions with the same final state particles but without forming the intermediate 
$(^4$He$-\eta)_{bound}$ state and, therefore, resulting on average in a much higher
c.m. momenta of $^3\mbox{He}$ (see Fig.~\ref{pmomc.m._p}).

The process described above  should result in a resonance-like structure in the excitation function of the $dd \rightarrow ^3$He$ p \pi^{-}$ and the $dd \rightarrow ^3$He$ n \pi^{0}$ reactions if we select events with low ${^3}$He center-of-mass (c.m.) momenta. 
According to the discussed scheme, there exist four equivalent decay channels
of the $({^4\mbox{He}}-\eta)_{bound}$ state: $({^4\mbox{He}}-\eta)_{bound} \rightarrow {^3\mbox{He}} p \pi^{-}$, $({^4\mbox{He}}-\eta)_{bound} \rightarrow {^3\mbox{He}} n \pi^{0}$, $({^4\mbox{He}}-\eta)_{bound} \rightarrow {^3\mbox{H}} p \pi^{0}$, $({^4\mbox{He}}-\eta)_{bound} \rightarrow {^3\mbox{H}} n \pi^{+}$.
In our experiment we concentrated on the first two out of the listed decay modes.

The first dedicated experiment with \wasa~ was performed in June 2008 by measuring the excitation fun\-ction of the $dd \rightarrow$ $^{3}\hspace{-0.03cm}\mbox{He} p \pi^{-}$  reaction near the $\eta$ meson production threshold.
The  analysis exhibited no structure which could be interpreted
as a resonance originating from the decay of the $\eta$-mesic ${^{4}\mbox{He}}$~\cite{Adlarson2013}. 
The upper limit for the formation and decay of the bound state in the process
$dd \rightarrow ({{^4\mbox{He}}-\eta})_{bound} \rightarrow {^3\mbox{He}} p \pi^{-}$ 
at the 90\% confidence level, was determined from 20~nb to 27~nb for the bound state width ranging from 5~MeV to 35~MeV, respectively. 

The achieved upper limit is only few times larger than recently predicted values~\cite{Wycech-Acta}.
Therefore there are chances for the observation of such a state from the 20 times higher statistics data collected in 2010 with \wasa.

\section{New experiment}

During the second experiment, in November 2010, two channels of the $\eta$-mesic helium decay were searched for:  $dd\rightarrow(^{4}\mbox{He}$-$\eta)_{bound}\rightarrow$ $^{3}\mbox{He} p \pi{}^{-}$ and  $dd\rightarrow(^{4}\mbox{He}$-$\eta)_{bound}\rightarrow$ $^{3}\mbox{He} n \pi{}^{0} \rightarrow$ 
${^{3}\mbox{He}} n \gamma \gamma$~\cite{Magda2}. 
During the experimental run the momentum of the deuteron beam was varied continuously within each acceleration cycle
from  2.127~GeV/c to 2.422~GeV/c, crossing the kinematic threshold for $\eta$ production in the $dd \rightarrow ^4$He$\,\eta$ reaction at 2.336~GeV/c.
This range of beam momenta corresponds to a variation of the $^4\mbox{He}$-$\eta$ excess energy  from -70~MeV to 30~MeV.

Data were taken for about~155 hours. 
Taking into account the fact that two reactions were measured, in total more than 20 times higher statistics were collected than in the 2008 run. 

\begin{figure}[htb]
  \centering
  \includegraphics[width=5.5cm,height=5cm]{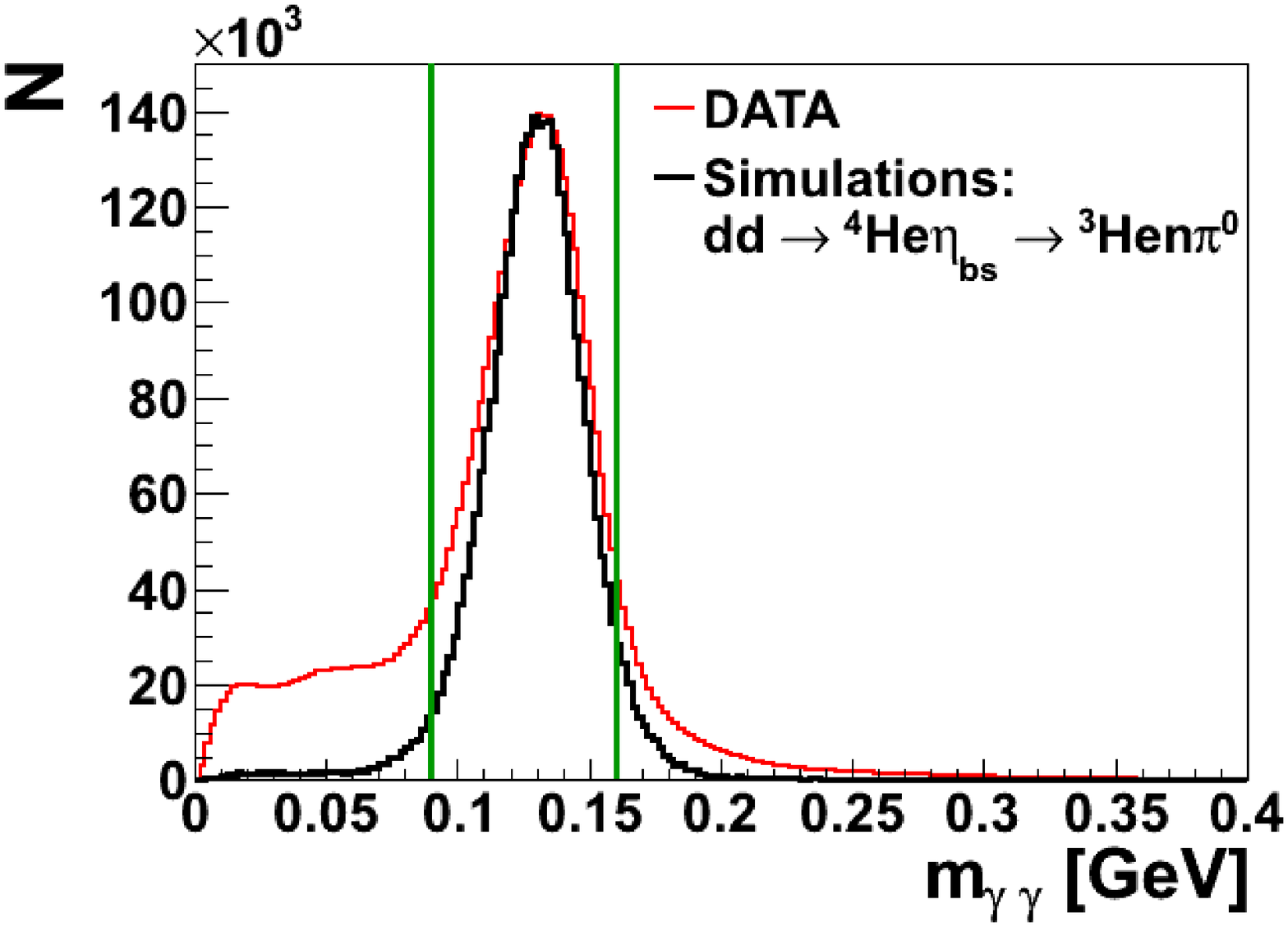} \includegraphics[width=5.5cm,height=5cm]{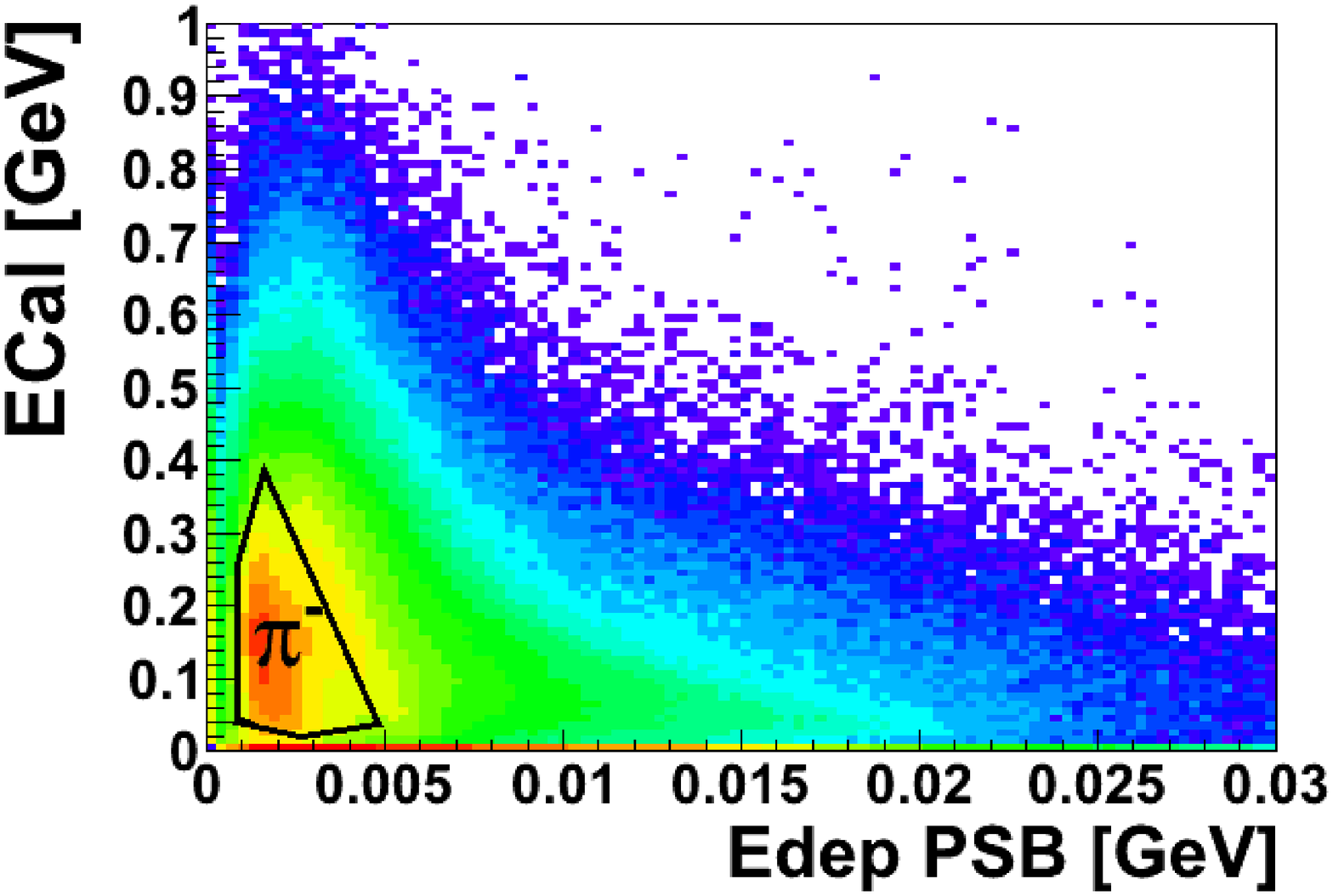}\\
  \vspace{-0.3cm}
  \caption{(Left) Distribution of the invariant mass of two gamma quanta. The vertical lines indicate the conditions applied to select pion candidates. (Right) Experimental distribution of the energy loss in the Plastic Scintillator Barrel ($x$-axis) combined with the energy deposited in the Electromagnetic Calorimeter ($y$-axis). The lines indicate the graphical condition applied to select pion candidates. \label{fig3}}
\end{figure}

Independent analyses for the $dd\rightarrow$ $^{3}\hspace{-0.03cm}\mbox{He} n \pi{}^{0}\rightarrow$ $^{3}\hspace{-0.03cm}\mbox{He} n \gamma \gamma$ and $dd\rightarrow$ $^{3}\hspace{-0.03cm}\mbox{He} p \pi{}^{-}$ reactions were carried out. The $^{3}\hspace{-0.03cm}\mbox{He}$ for both cases was identified in the Forward Detector based on the \mbox{$\Delta$E-E method}. The $\pi^{0}$ was reconstructed in the Central Detector from the invariant mass of two decay photons while the $\pi^{-}$ identification in Central Detector was based on the measurement of the energy loss in the Plastic Scintillator Barrel combined with the energy deposited in the Electromagnetic Calorimeter. Neutrons and protons were identified via the missing mass technique. The corresponding spectra with applied cuts are presented in Fig.~\ref{fig3}.

\begin{figure}[htb]
  \centering
  \includegraphics[width=5.5cm,height=5cm]{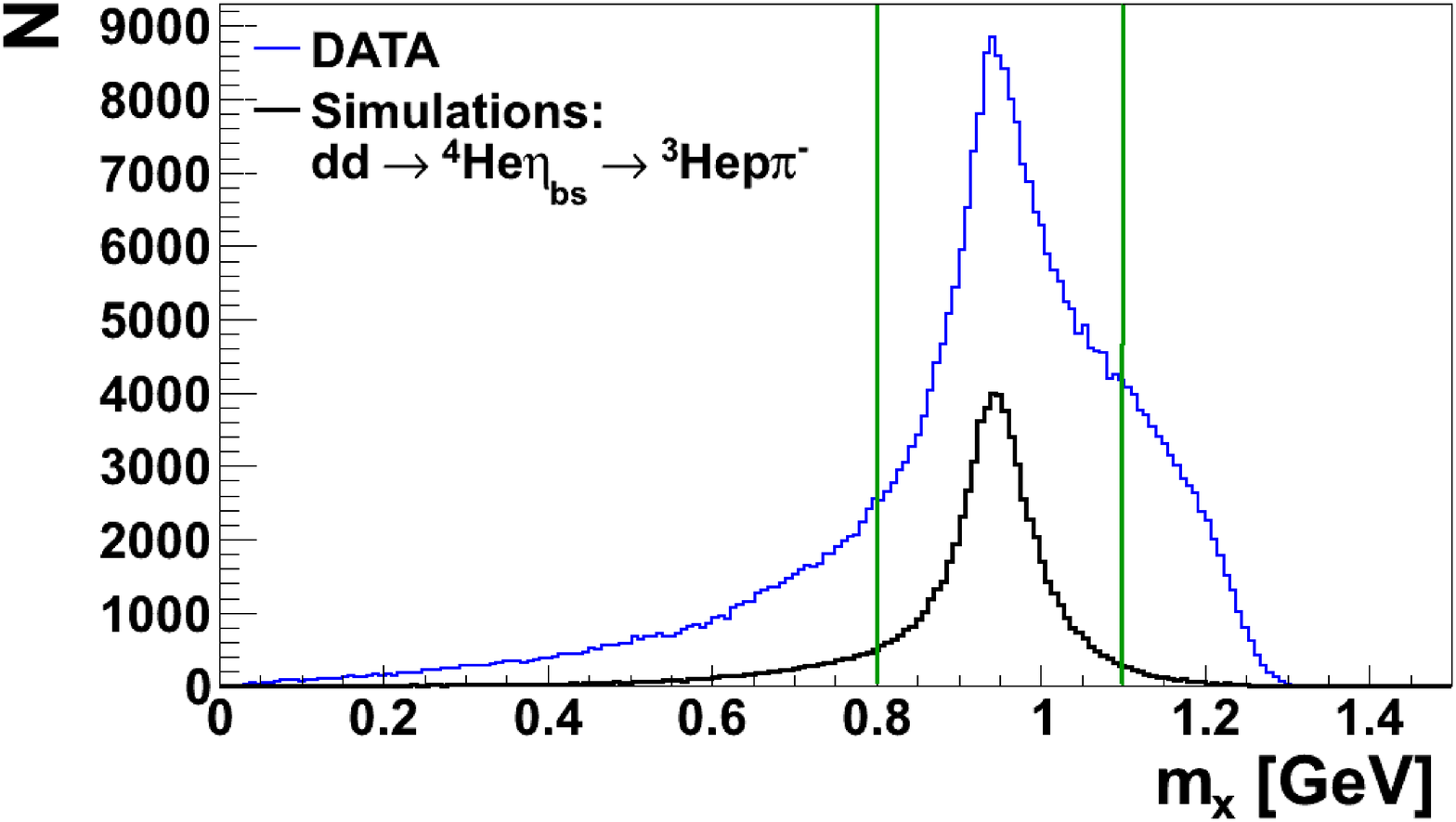} \includegraphics[width=5.5cm,height=5cm]{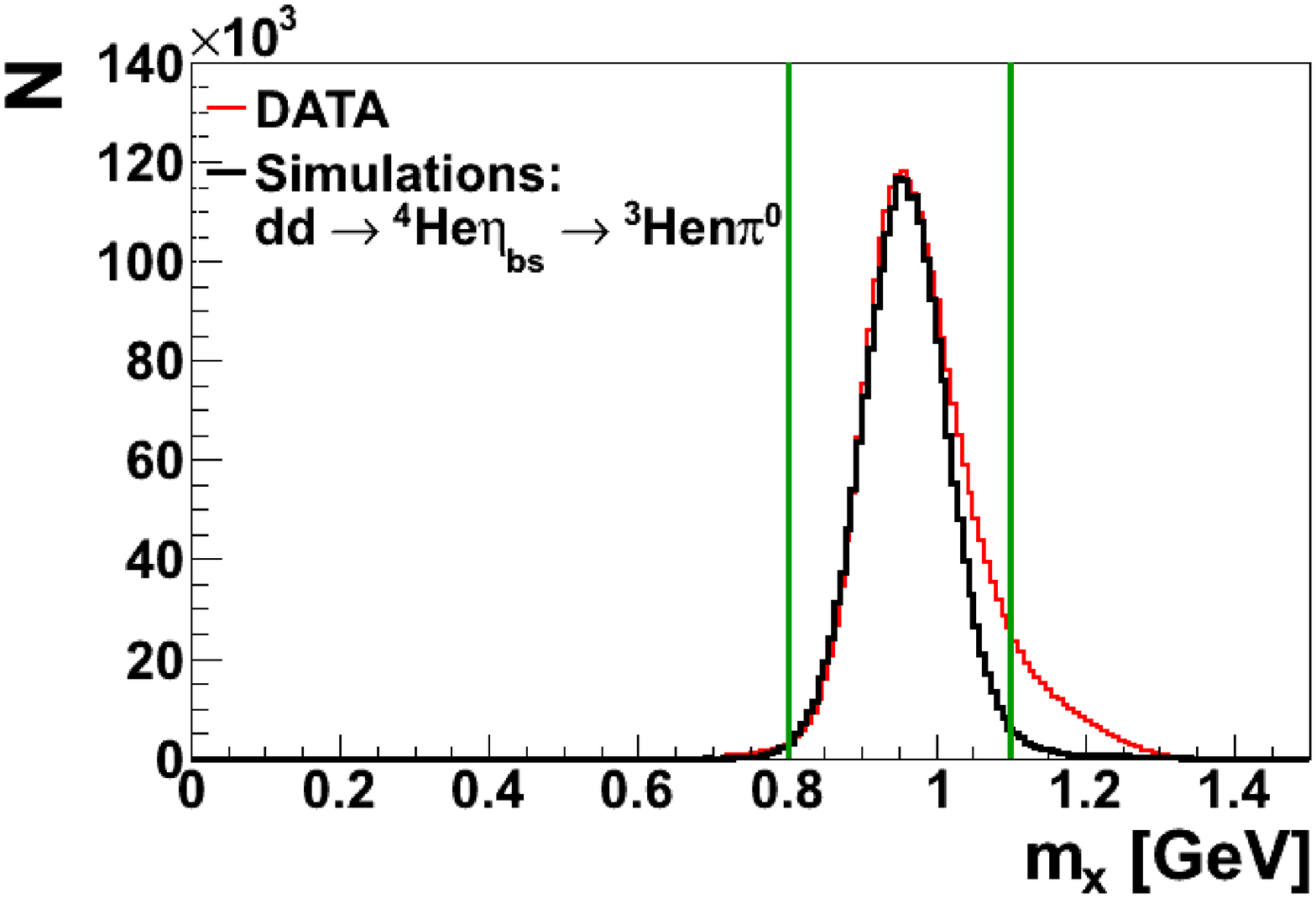}
  \vspace{-0.3cm}
  \caption{ Proton and neutron identification. The data are shown as red and blue histograms, the Monte Carlo simulations of the signal are marked with the black line, while the applied cuts are marked in green. $m_{x}$ denotes the missing mass for the $dd\rightarrow$ $^{3}\hspace{-0.03cm}\mbox{He} X$ reaction.~\label{fig4}}
\end{figure}

\begin{figure}[htb]
\centering
\includegraphics[width=5.5cm,height=5cm]{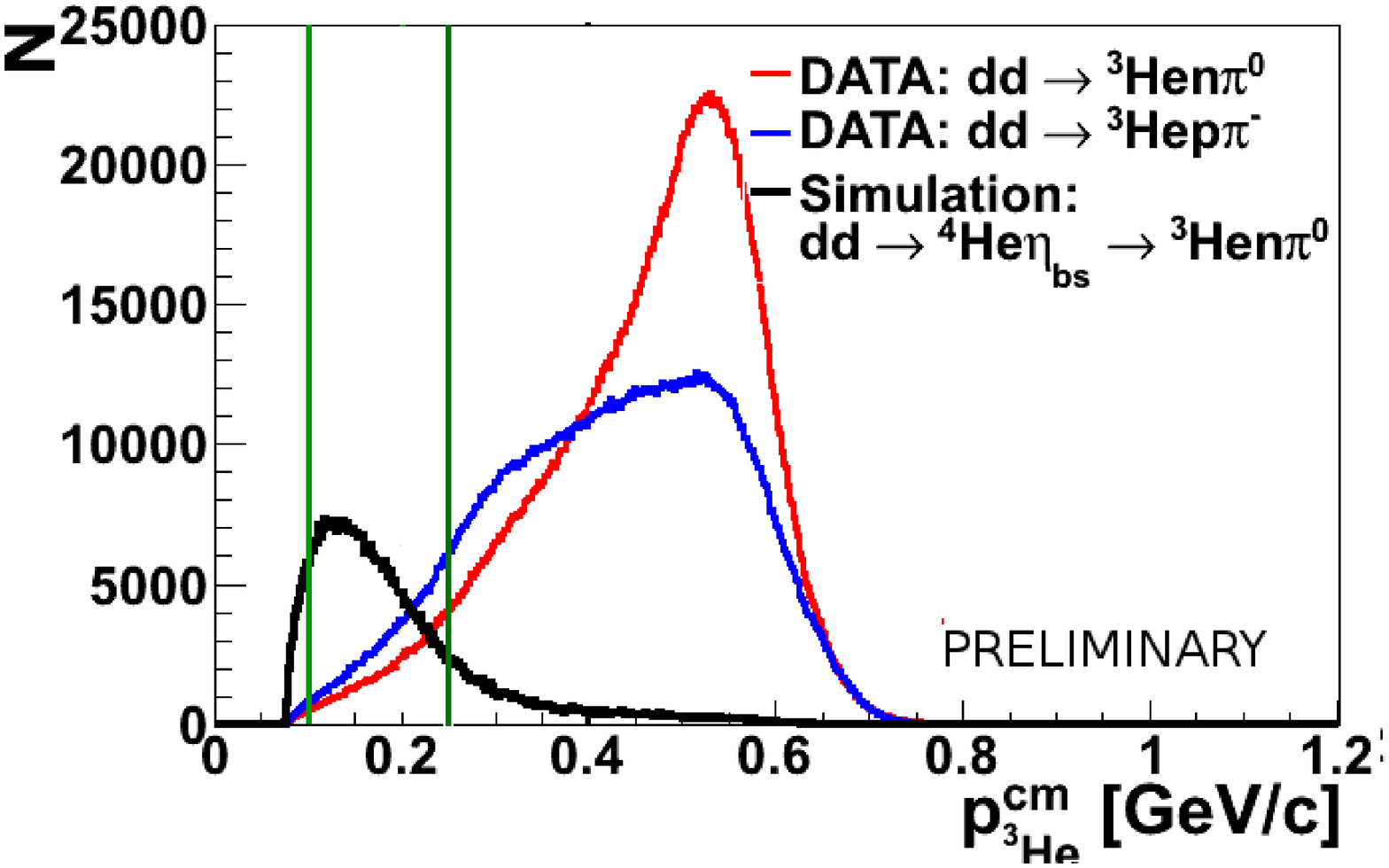} 
\includegraphics[width=5.5cm,height=5cm]{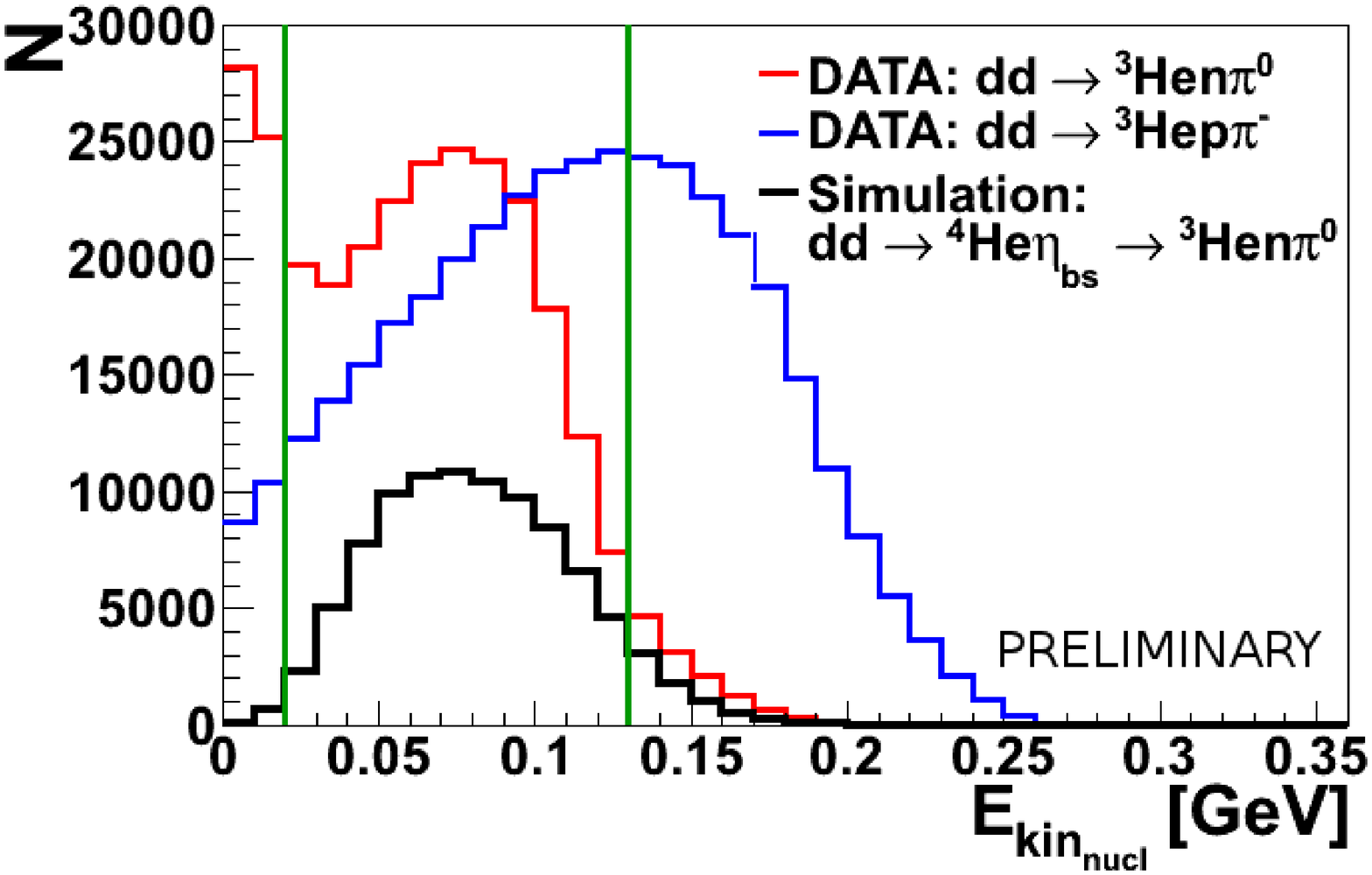}
\caption{Spectrum of $p^{cm}_{^{3}\hspace{-0.05cm}He}$ (left panel), $E^{cm}_{kin_{nucl}}$ (right panel). Data are shown in red and blue for $dd\rightarrow$ $^{3}\hspace{-0.03cm}\mbox{He} n \pi{}^{0}$ and $dd\rightarrow$ $^{3}\hspace{-0.03cm}\mbox{He} p \pi{}^{-}$ reaction, respectively. Monte Carlo simulations of the signal are shown in black, while the applied cuts are marked with the green lines.~\label{fig5}}
\end{figure}

\begin{figure}[htb]
\centering
\includegraphics[width=5.5cm,height=5cm]{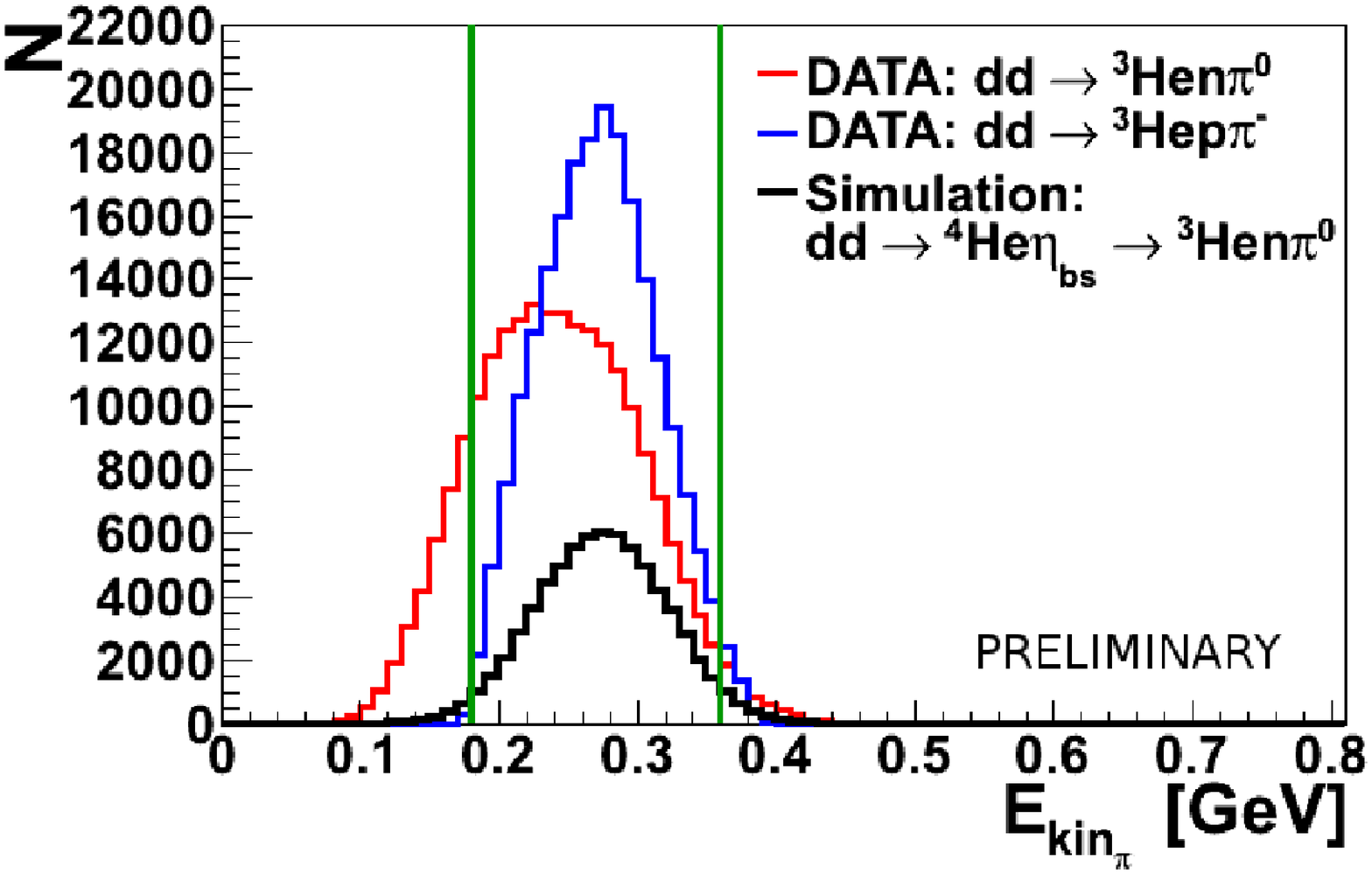} 
\includegraphics[width=5.5cm,height=5cm]{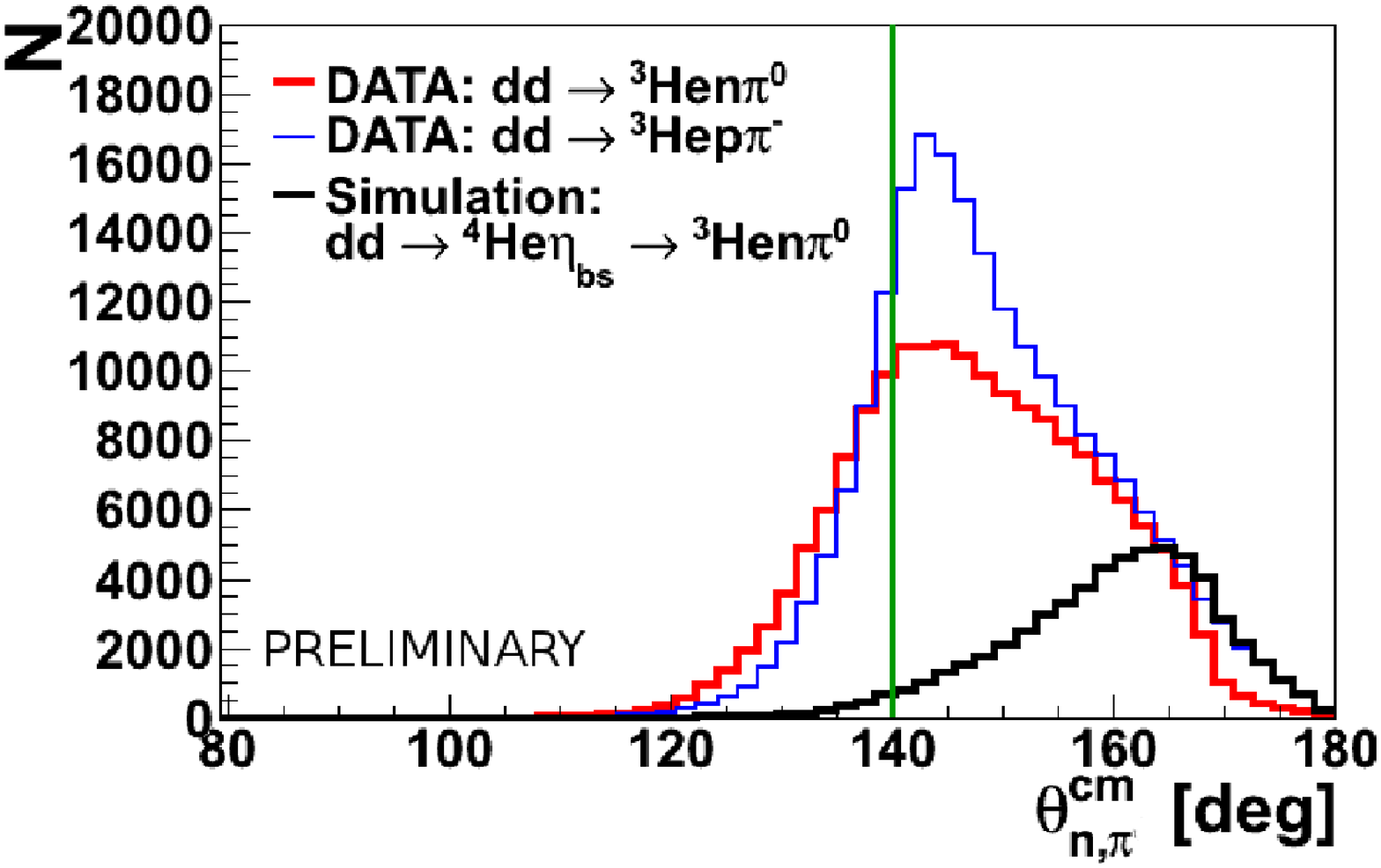}
\caption{Spectrum of $E^{cm}_{kin_{\pi}}$ (left panel) and $\theta^{cm}_{nucl,\pi}$ (right panel). Data are shown in red and blue for $dd\rightarrow$ $^{3}\hspace{-0.03cm}\mbox{He} n \pi{}^{0}$ and $dd\rightarrow$ $^{3}\hspace{-0.03cm}\mbox{He} p \pi{}^{-}$ reaction, respectively. Monte Carlo simulations of the signal are shown in black, while the applied cuts are marked with the green lines.~\label{fig6}}
\end{figure}

In order to select events corresponding to the production of bound states, additional cuts in the $^{3}\hspace{-0.03cm}\mbox{He}$ c.m.  momentum, nucleon c.m. kinetic energy, pion c.m. kinetic energy and the opening angle between nucleon-pion pair in the c.m. were applied based on Monte Carlo simulations. These cuts are presented in Fig.~\ref{fig5} and in Fig.~\ref{fig6}.

\section{Preliminary results}

\begin{figure}[htb]
\includegraphics[width=5.5cm,height=5cm]{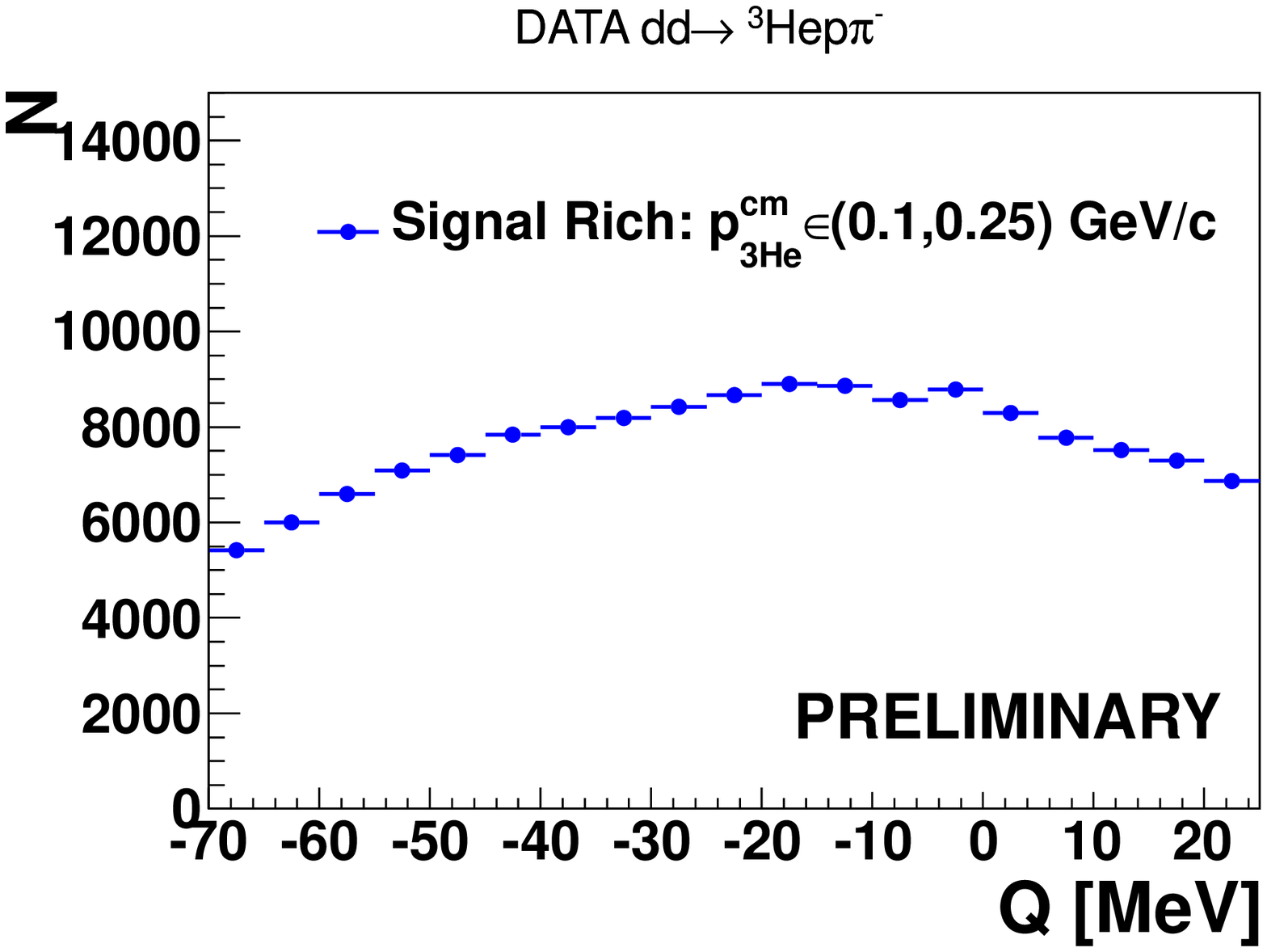}
\includegraphics[width=5.5cm,height=5cm]{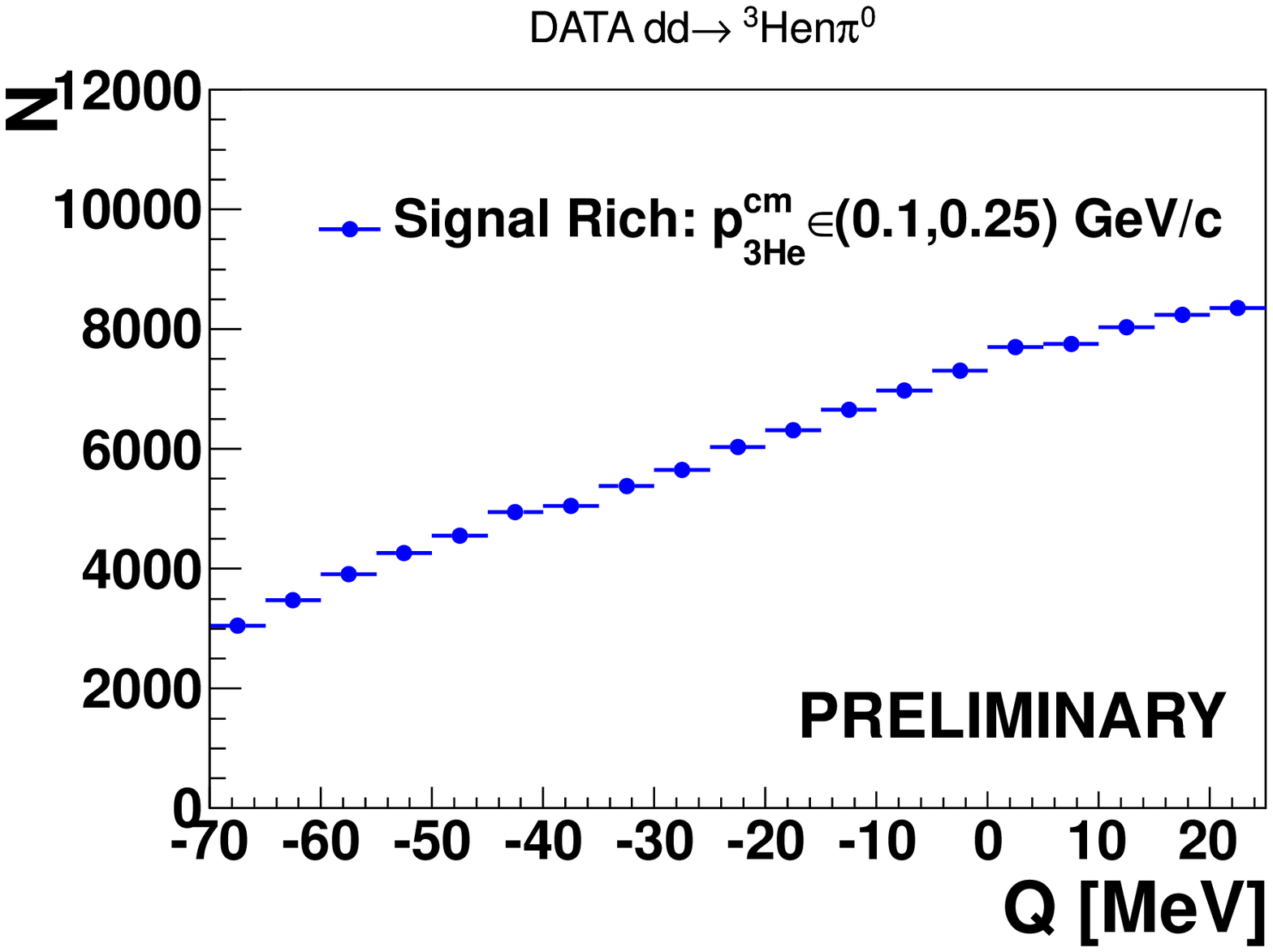}
\caption{\label{fig:excit2}Preliminary excitation function for the $dd \rightarrow {^3\mbox{He}} p \pi^-$  and  for the $dd \rightarrow {^3\mbox{He}} n \pi^{0}$ reactions under condition that the ${^3\mbox{He}}$ momentum in c.m. frame is in the range from 0.1 to 0.25 GeV/c ("signal-rich" area). The distributions are not corrected for efficiency. }

The difference in the shape of the presented spectra could be explained by the background contributions from different isospin combinations. Studies on a more detailed background description are in progress.

\end{figure}

We present the preliminary excitation function for the "signal-rich" region, corresponding to the low ${^3\mbox{He}}$ momenta in c.m. frame. 
The  region corresponds to the momenta in the range from 0.1 to 0.25 GeV/c (Fig.~\ref{fig:excit2}).

\section{Conclusions}

We perform the search of the $\eta$-mesic helium with \wasa~ via deuteron-deuteron collision.
We concentrated on the decay model where the $\eta$ is absorbed on one of the nucleons 
which subsequently decays into $N-\pi$ pair.  
The preliminary analysis of the uncorrected excitation functions for the "signal-rich" regions, 
does not exhibit any sharp structure.
We are continuing the investigations, and as a next step, we will compare the excitation function in the 
different ${^3\mbox{He}}$ momentum regions. The collected statistics are sufficient to observe a difference in the shapes if the cross section for the production and decay of the bound state in the   $dd \rightarrow {^3\mbox{He}} p \pi^-$  and $dd \rightarrow {^3\mbox{He}} n \pi^{0} \rightarrow$ $^{3}\hspace{-0.03cm}\mbox{He} n \gamma \gamma$ reactions would be of the order of the predictions in~\cite{Wycech-Acta}.

\section{Acknowledgments}
This work was supported by the the Polish National Science Center
under grant and No. 2011/01/B/ST2/00431.


\end{document}